\pdfoutput=1
\documentclass[runningheads]{llncs}
\usepackage[T1]{fontenc}
\usepackage{booktabs}
\usepackage{multirow}
\usepackage{amsmath}

\usepackage{graphicx}

\begin{document}

\title{Training Text-to-Speech Model with Purely Synthetic Data:
Feasibility, Sensitivity, and Generalization Capability}

\titlerunning{Training Text-to-Speech Model with Purely Synthetic Data}

\author{Tingxiao Zhou \and
Leying Zhang \and Zhengyang Chen \and
Yanmin Qian}

\authorrunning{Z. Tingxiao et al.}

\institute{Auditory Cognition and Computational Acoustics Lab \\
MoE Key Lab of Artificial Intelligence, AI Institute \\ 
School of Computer Science, Shanghai Jiao Tong University, Shanghai, China
\email{ztxiao1211, zhangleying, zhengyang.chen, yanminqian@sjtu.edu.cn} 
}

\maketitle             

\begin{abstract}
The potential of synthetic data in text-to-speech (TTS) model training has gained increasing attention, yet its rationality and effectiveness require systematic validation. In this study, we systematically investigate the feasibility of using purely synthetic data for TTS training and explore how various factors—including text richness, speaker diversity, noise levels, and speaking styles—affect model performance. Our experiments reveal that increasing speaker and text diversity significantly enhances synthesis quality and robustness. Cleaner training data with minimal noise further improves performance. Moreover, we find that standard speaking styles facilitate more effective model learning. Our experiments indicate that models trained on synthetic data have great potential to outperform those trained on real data under similar conditions, due to the absence of real-world imperfections and noise.
\end{abstract}


\section{Introduction}
Text-to-Speech (TTS) systems refer to technologies that automatically convert written text into intelligible and natural-sounding spoken audio~\cite{tan2021survey,lengprompttts,zhang2024covomix}. Among various TTS paradigms, zero-shot TTS has recently gained significant attention due to its capability to synthesize speech for previously unseen speakers by leveraging only a single speech prompt or a short reference utterance. This ability to generalize to new speakers without requiring speaker-specific training data represents a substantial advancement in speech synthesis. Recent advances in deep learning have significantly enhanced the quality of synthesized speech, making zero-shot TTS increasingly popular for generating natural and expressive speech that closely approximates human-level naturalness~\cite{wang2023neural,le2024voicebox,casanova2024xtts,zhang2025covomix2,10800670,liu2025e2e,guo2025splitmeanflow}. 

As the scaling law demonstrates significant effectiveness in many machine learning fields~\cite{kaplan2020scaling,li2025less,zhang2024beyond}, the development of high-quality text-to-speech  heavily relies on large-scale, high-fidelity speech datasets~\cite{lajszczak2024base,anastassiou2024seed}. However, accumulating a large scale dataset is challenging in three main aspects. First, collecting and annotating such data is both costly and time-consuming. Second, real-world speech data often comes with privacy concerns~\cite{bellovin2019privacy}. Third, real-world datasets are inherently noisy, containing recording imperfections and environmental distortions that can degrade model performance~\cite{zhang2025noisepreserve,zhang2021denoispeech}.

To address these challenges, synthetic data offers a promising alternative by eliminating the need for human-recorded speech while ensuring broad availability without ethical or legal constraints. The advanced TTS models and technologies have been widely used for augmenting the training data in many speech-related tasks such as automatic speech recognition~\cite{fazel2021synthasr,eigenschink2023deep,rossenbach2024effect}, speaker verification~\cite{du2021synaug} and speech enhancement~\cite{li2025less,zhang2024scale,zhang2024ddtse}. Their scaling effects of each data attribute are also investigated by using purely synthetic data in training. Several studies also explored the use of synthetic data in TTS training to solve the cross-style transfer and accent transfer problems~\cite{ueda2024exploring,finkelstein2022training}.  However, there is limited research that systematically examines why synthetic data is valuable for training TTS models and which data attributes contribute most to their effectiveness. 

In this study, by leveraging clean, noise-free synthetic data, we aim to reveal the feasibility, sensitivity and generalization capability of incorporating  synthetic data into TTS model training. The contribution of this paper can be summarized in three major points as follows:
    
1) We explore the feasibility of training TTS models purely with synthetic data and demonstrate that the resulting models have great potential to achieve better performance than models trained on real data.

2) We conduct a comprehensive analysis of the sensitivity of model performance to key data factors, including text richness, speaker diversity, noise level, and speaking style.

3) We assess the generalization capability of the models by evaluating their performance when transferring from reading-style training data to out-of-domain casual data.

\section{Related Work}
\subsection{Large Scale Text-to-Speech Model}
In recent years, large-scale zero-shot TTS models have significantly advanced, achieving remarkable improvements in speech naturalness and expressiveness. Diffusion or Flow-matching based Zero-Shot TTS~\cite{le2024voicebox,chen2024f5} is one of the representative methods, which handle the problem in a non-auto regressive manner. Auto-regressive TTS~\cite{casanova2024xtts,ChatTTS2024}, especially neural-codec based model~\cite{wang2023neural,borsos2023soundstorm,zhang2023speak,wang2024speechx}, on the other hand, predicts each audio sample or feature frame sequentially. Both methods significantly improve the quality and versatility of synthesized speech, demonstrating the role of scaling laws in the TTS task. In this work, we choose several SOTA or influential large-scale TTS models, including XTTS-v2, CosyVoice2, and ChatTTS as the pre-trained model to synthesize training datasets for experiment.

XTTS-v2\footnote{https://huggingface.co/coqui/XTTS-v2}~\cite{casanova2024xtts} is an auto-regressive text-to-speech model that generates audio by predicting discrete audio tokens produced by a pre-trained discrete Variational Autoencoder (VAE). It is trained with 16k hours of data mostly consisting of public datasets. The entire training process consists of approximately 1 million steps. and the final model comprises approximately 750 M parameters.

CosyVoice2\footnote{https://github.com/FunAudioLLM/CosyVoice}~\cite{du2024cosyvoice} is a multilingual speech synthesis
model based on supervised discrete speech tokens and a pre-trained LLM. A 200,000-hour dataset is used to train the speech tokenizer with normalized transcriptions as labels. And the model is trained on a dataset approximately 150,000 hours. The open-source version on HuggingFace comprises approximately 0.5 B parameters, which is used to synthesize speeches with high naturalness and quality.

ChatTTS\footnote{https://github.com/2noise/ChatTTS.git}~\cite{ChatTTS2024} is a text-to-speech model designed specifically for dialogue scenarios such as LLM assistant. It uses the Llama model to predict the audio tokens for the masked text parts. The open-source version on HuggingFace is a 40,000 hours pre-trained model without SFT, which is used to synthesize oral-style and conversational speech.

\subsection{Efficient Text-to-Speech Models}
Efficient Text-to-Speech (TTS) models are designed to generate high-quality, speech while minimizing computational resources and latency. Traditional autoregressive TTS models generate speech sequentially, which can lead to slower inference times. To overcome this, non-autoregressive models like FastSpeech2~\cite{ren2020fastspeech} enables parallel generation of speech frames and significantly reduces latency. Recent TTS models~\cite{le2024voicebox,zhang2024covomix,zhang2025noisepreserve} have demonstrated the capability of flow-matching models to generate high-quality speeches. 

Matcha-TTS\footnote{https://github.com/shivammehta25/Matcha-TTS}~\cite{mehta2024matcha}, is a popular flow-matching TTS model and the codebase of SOTA model, CosyVoice2. It utilizes a non-autoregressive encoder-decoder architecture to achieve fast and lightweight speech synthesis. The main architecture is a U-Net containing 1D convolutional residual blocks to downsample and upsample the inputs. The total params is approximately 20.9 M. In the following experiments, we use Matcha-TTS as our TTS model backbone.

\subsection{Knowledge Distillation}
Knowledge distillation (KD) is a wildly used machine learning method~\cite{hinton2015distilling} that can transfer the robust representation of a teacher model to student models~\cite{zhang2021knowledge}. RKD~\cite{cheng2024residual} utilizes the residual knowledge distillation from multiple teacher layers to reduce the computational burden. TTS-by-TTS~\cite{9414408} uses high-quality TTS system to improve the quality of target TTS system. However, less attention has been paid to knowledge distillation for purely
synthetic data level.

As shown in Table~\ref{tab:model}, Matcha-TTS is a much smaller model than teacher models in terms of parameter and training dataset. In the following experiments, we use Matcha-TTS as our TTS model backbone, and we want to leverage the large model’s strength to help improving the smaller model’s performance, focusing on how to achieve better KD results at the data level.

\begin{table}[t]
    \centering
    \caption{\textbf{Comparison of Model Scale.} “Teacher” refers to large scale TTS models used to generated synthetic training datasets. “Student” refers to relatively small model used for training.}
    \label{tab:model}
    \begin{tabular}{c|c|cc}
    \toprule
        Role & Model & Parameter & Training Data  \\
         \midrule
        \multirow{3}{*}{Teacher} & XTTS-v2 & 750 M & 16k h \\
        & CosyVoice2 & 0.5 B & 200k h \\
        & ChatTTS & 3 B & 40k h \\
         \midrule
        Student & Matcha-TTS & 20.9 M & 100 h (Our Setting) \\
    \bottomrule
    \end{tabular}
    % \vspace{-0.35cm}
\end{table}

\begin{table*}[t]
    \centering
    \caption{\textbf{Performance Comparison on LibriSpeech and SeedTTS test sets.} “Ground Truth” means calculating evaluation metrics (skip SIM) directly on the two test dataset. “XTTS-v2”, “CosyVoice2”, and “ChatTTS” mean using the three large scale TTS models to infer test dataset respectively. Since ChatTTS does not support zero-shot TTS tasks, there is no SIM metrics in this case. “Matcha-TTS” means our experiment on four different training data. $\mathcal{R}$ refers to real data LibriSpeech train-100. $\mathcal{F}_{XTTS-v2}$, $\mathcal{F}_{CosyVoice2}$, and $\mathcal{F}_{ChatTTS}$ refer to synthetic data generated by the three large scale TTS models respectively.}
    \label{tab:feasibility}
    \resizebox{\columnwidth}{!}{
    \begin{tabular}{c|c|cccc|cccc}
    \toprule
         \multirow{2}{*}{Model} & \multirow{2}{*}{Training Data} & \multicolumn{4}{c|}{LibriSpeech test-clean} & \multicolumn{4}{c}{Seed-TTS test} \\ 
         ~ & ~ &WER $\downarrow$& SIM $\uparrow$ & DNSMOS $\uparrow$& UTMOS$\uparrow$ & WER $\downarrow$& SIM$\uparrow$  & DNSMOS$\uparrow$ & UTMOS $\uparrow$ \\ \midrule
        Ground Truth & / & 5.395 & / &  3.84 & 4.08 & 4.413 & / & 	3.77 &	3.53   \\ \midrule
        XTTS-v2 & / & 4.639&	0.544&		3.80 &3.73 &3.627 &	0.462  &3.79  &	3.54   \\ 
       CosyVoice2  & /  & 4.070 & 0.694 & 3.87  & 4.36   &  2.493	&0.663&	3.86  &	4.09  \\ 
        ChatTTS & / & 9.432&	/		&3.83 &3.55&	5.592&	/	&	3.78 &3.67  \\ 
        \midrule
      \multirow{4}{*}{Matcha-TTS} & $\mathcal{R}$ & 4.428 & 0.389 & 	 3.71  & 3.74  & 3.595 & 0.342 &	3.76& 3.78 \\ 
    ~ & $\mathcal{F}_{XTTS-v2}$  & 3.493&	0.404&	3.72 &	3.75 &	3.368&	0.338&	3.72 &3.66   \\ 
    ~ & $\mathcal{F}_{CosyVoice2}$  & 4.990 &	0.383& 3.80 & 3.83 &	4.184&	0.334&	3.85 &	3.82   \\ 
      ~ & $\mathcal{F}_{ChatTTS}$   & 9.007	&0.276&		3.68&	3.18 &	11.431&	0.282&	3.68 &	3.21 \\ 
    \bottomrule
    \end{tabular}
    % \vspace{-0.35cm}
    % 10.428  2.470
    % 0.796
    }
\end{table*}

\section{Methodology and Experimental Design}
To systematically investigate the utilization of purely synthetic data in TTS model training and to assess model sensitivity to various data attributes—such as text richness, speaker diversity, noise level, and speaking style—we employ a generation–training–evaluation framework. 

For each attribute, we first generate a corresponding dataset tailored to that attribute. We then train a TTS model using this dataset. Finally, we evaluate the trained model on two distinct datasets using a comprehensive set of metrics to quantify its performance under different conditions.

\subsection{Generation}
To analyze the impact of each attribute in the speech data, we generate a series of purely synthetic training datasets by utilizing a pre-trained zero-shot TTS model (i.e., XTTS-v2), which can precisely and independently control the textual content and speaker identity via prompting. All generated datasets have the same number of utterances (denoted  $m$) as the real dataset $\mathcal{R}$.

Firstly, to explore the feasibility of purely synthetic data for training a TTS model, we generate a synthetic dataset $\mathcal{F}$ which consist the same number of text and speaker as $\mathcal{R}$, except that it is totally synthesized. We completely shuffle the speaker and text pair while generating the synthetic data, ensuring the experiment data unseen by the pre-trained zero-shot TTS model.

Then, we explore the sensitivity of model performance to four key data factors. Specifically, we analyze  the influence of speaker diversity, text richness, noise level, and speaking style by manipulating the number of unique speakers, the content of different texts, the intensity and proportion of noise, and the type of pre-trained model respectively.

1) To manipulate the \textbf{speaker diversity}, we generate multiple datasets $\mathcal{S}_s$ with randomly selected $s$ speakers. One utterance of each speaker is chosen as the speaker prompt, which will be reused $\frac{m}{s}$ times. 

2) To investigate the influence of \textbf{text richness}, we generate multiple datasets $\mathcal{T}_p$ with the reuse rate $p$. Before generation, we replace $p*m$ utterance transcriptions in $\mathcal{R}$ with repeated texts, and we strictly constrain the total number of words between $\mathcal{R}$ and $\mathcal{T}_p$ to be the same to roughly retain the overall duration. Specifically, $\mathcal{T}_p$ has $m$ utterances with $m*(1-p)$ different transcriptions. Each utterance in $\mathcal{T}_0$  has a unique text, whereas the content of all utterances in $\mathcal{T}_1$  is identical. 

3) To analyze the influence of \textbf{noise level}, with a given signal to noise ratio (SNR) and a given noise ratio $p_n$, we randomly sample noises from the noise dataset $\mathcal{N}$ and mix with the  $m*p_n$ utterances from $\mathcal{F}$. We also augment the clean dataset $\mathcal{F}$ with reverberation from the  dataset $\mathcal{V}$ with a given ratio $p_r$.

4) To analyze the influence of \textbf{speaking style}, we select three large-scale TTS models—XTTS-v2, CosyVoice2, and ChatTTS — to generate data of reading-style, natural-style and oral-style speech respectively, while keeping all other attributes constant.

\subsection{Training}
We use LibriSpeech~\cite{panayotov2015librispeech} dataset train-100 set as the real speech dataset $\mathcal{R}$, which contains about 100 hours of clean speech. It needs to be clarified that the LibriSpeech train-100 set contains only 251 speakers. In our experiment, we add additional speakers from train-360 set,  involving up to 1,000 speakers. We utilize WHAM!~\cite{wichern2019wham} dataset as the noise dataset $\mathcal{N}$, and RIRS~\cite{ko2017study} dataset as the reverberation dataset $\mathcal{V}$. To enable Matcha-TTS to handle zero-shot tasks, we additionally use the WeSpeaker~\cite{wang2022wespeakerresearchproductionoriented} pre-trained ResNet221\_LM\footnote{https://github.com/wenet-e2e/wespeaker.git} model to extract speaker embeddings and replace the original speaker id. We train Matcha-TTS for 100 epochs with a batch size of 32 on the synthetic training datasets. We set the learning rate to 1e-4. We utilize the Adam optimizer and 2 Nvidia A10 GPU for training. We fix the hyperparameters for all experiments.

\subsection{Evaluation}
We use the official LibriSpeech test-clean  set for in-domain evaluation, and the  Seed-TTS~\cite{anastassiou2024seed} test set for out-of-domain evaluation. We adopt four metrics for evaluation. 
We calculate word error rate (WER) with Whisper-large-v3~\cite{radford2023robust} as the automatic speech recognition engines. We measure the speaker similarity (SIM) by calculating cosine distance of speaker embeddings of speech samples and their references, extracted from WavLM-large fine-tuned~\cite{chen2022wavlm}. We use DNSMOS~\cite{reddy2021dnsmos} and UTMOS~\cite{saeki2022utmosutokyosarulabvoicemoschallenge} for evaluating the generated speech clarity and overall quality. SIM, DNSMOS and UTMOS are the higher the better, while WER is the lower the better.

\section{Results and Analysis}
\subsection{Feasibility for Purely Synthetic data}
Table~\ref{tab:feasibility} shows the performance comparison of models trained on synthetic data and real data, and the large scale TTS model. Compared to XTTS-v2, a much smaller Matcha-TTS trained on its synthetic data achieves superior performance in terms of WER and UTMOS. However, the speaker similarity decreases due to the significant gap in training data size and model scale. 
Moreover, we observe that Matcha-TTS trained on the purely generated dataset (by XTTS-v2) performs better than those trained on real data. These findings validate our later investigations based on purely synthetic speech data. We will further analyze the potential reasons of this phenomenon in Section \ref{sec:noise}.

\subsection{Effects of Speaker Diversity}

\begin{figure*}[t]
  % \hspace{-0.5cm} 
  \includegraphics[width=\linewidth]{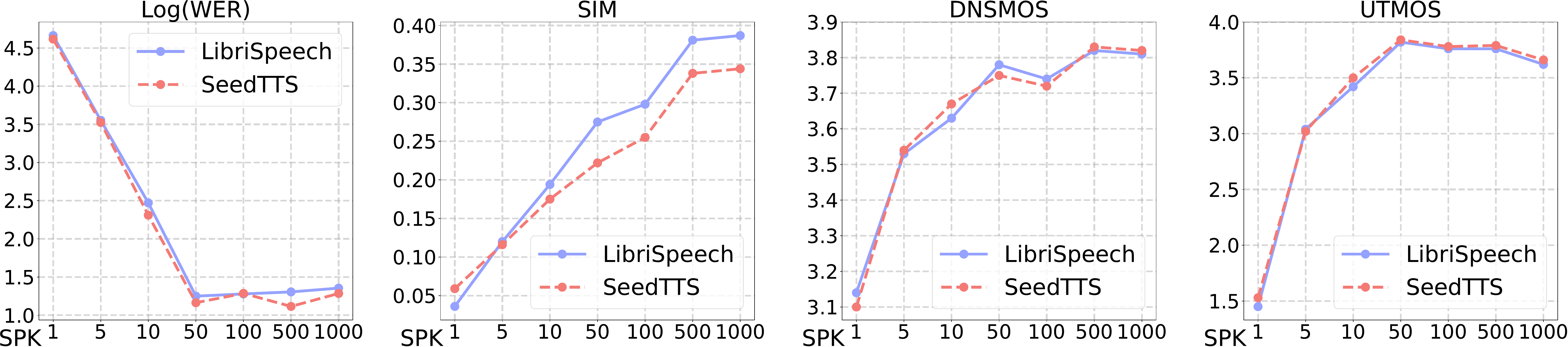}
  % \vspace{-0.45cm}
  \caption{Analysis of effects of the speaker diversity}
  \label{fig:spk}
  % %\vspace{-0.45cm}
\end{figure*}

\begin{figure*}[t]
  % \hspace{-0.4cm} 
  \includegraphics[width=\linewidth]{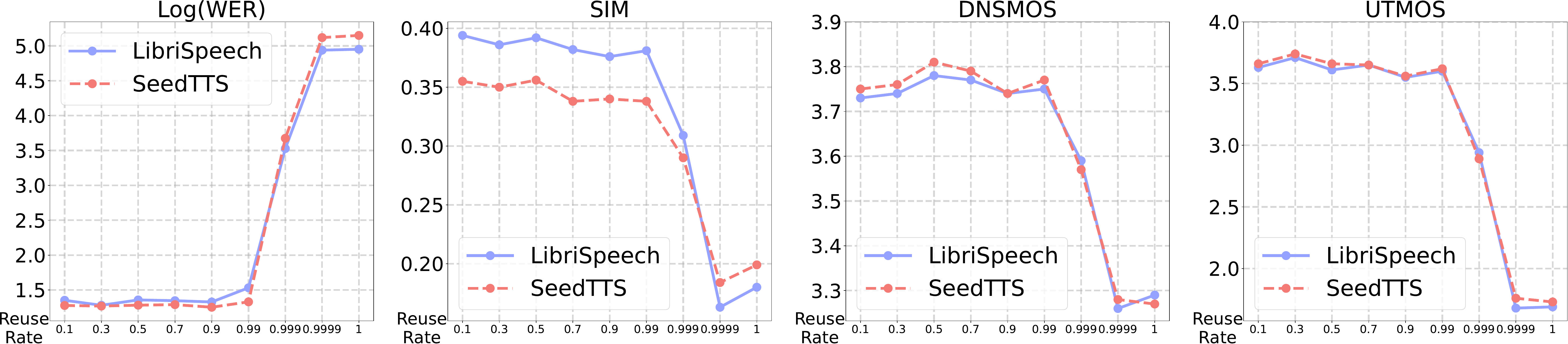}
  % \vspace{-0.45cm}
  \caption{Analysis of effects of the text richness}
  \label{fig:text}
  % %\vspace{-0.45cm}
\end{figure*}

Figure~\ref{fig:spk} illustrates the results on the impact of speaker diversity on model performance. The performance improves consistently as the number of speakers increases across all evaluation metrics, suggesting that greater speaker variability enhances the system’s generalization and robustness by exposing the model to a wider range of speaker characteristics.

However, when the number of speakers exceeds 50, improvements in Word Error Rate (WER) and Unweighted Task Mean Opinion Score (UTMOS) begin to plateau, indicating the presence of a performance bottleneck likely caused by limitations in data volume and model capacity. Interestingly, the turning point for the SIM metric occurs much later, at around 500 speakers, implying that speaker diversity plays a more critical role in accurately modeling speaker-specific characteristics than in capturing speech pronunciation.

Despite this, it is important to note that due to constraints in model size and the scale of training data, our model trained solely on synthetic data with 1,000 speakers still exhibits a significant gap in speaker similarity compared to larger, more advanced models such as XTTS-v2. This suggests that, beyond speaker diversity, model capacity and real data quality remain crucial factors for achieving high speaker similarity.

\subsection{Effects of Text Richness}
Figure~\ref{fig:text} presents the results of the text richness experiment, showing that with the fixed total data duration, increasing the text diversity leads to improved performance across all four  metrics. This trend highlights the importance of text richness of the training data, as more text provides the model with a richer linguistic context. However, a turning point is observed at the reuse rate of 99\% (only 285 distinct text left), and further increases in text quantity result in marginal gains. This suggests that after reaching a certain threshold, additional text does not significantly contribute to performance improvements. 

Since Matcha-TTS is phoneme-based TTS model, we further conduct phoneme statistical analysis for various text richness through G2P experiment. As shown in Table~\ref{tab:phoneme}, when text reuse rate is below 99\%, the existing texts already contain all the 69 phonemes in English. Therefore, in resource-limited scenarios, a dataset with repeated text containing enough phoneme required for the model to learn is still sufficient for effective training.  

\begin{table}[t]
    \centering
    \caption{\textbf{Phoneme Statistics for Various Text Richness.}}
    \label{tab:phoneme}
    \begin{tabular}{c|ccccc}
    \toprule
         Reuse Rate & $\le 90\%$ & $99\%$ & $99.9\%$ & $99.99\%$  & $100\%$  \\
         \midrule
        Phonemes & 69 & 68 & 60 & 47 & 39 \\
    \bottomrule
    \end{tabular}
    % \vspace{-0.35cm}
\end{table}

\begin{figure*}[t]
  % \hspace{-0.45cm} 
  \includegraphics[width=\linewidth]{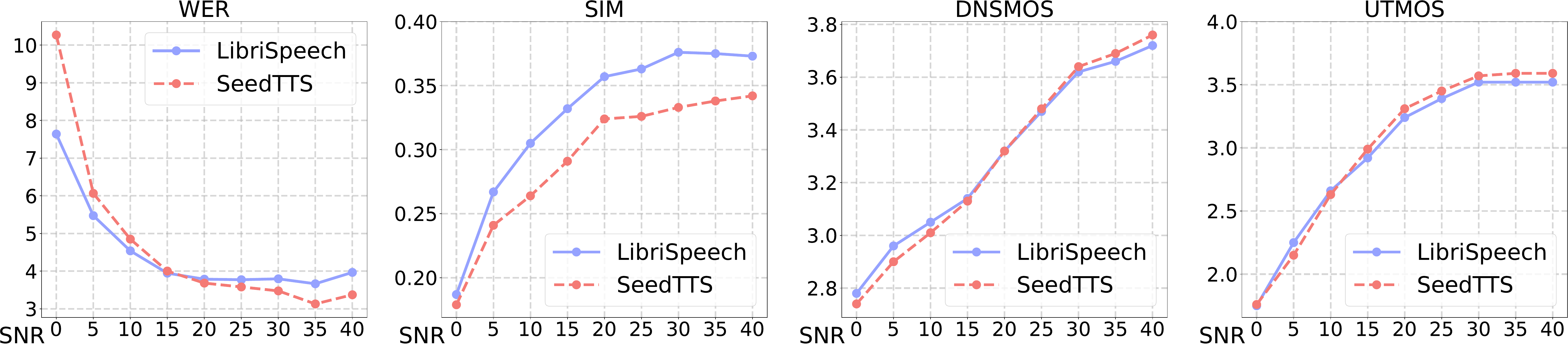}
  % \vspace{-0.45cm}
  \caption{Analysis of effects of the signal-to-noise ratio}
  \label{fig:snr}
  % %\vspace{-0.45cm}
\end{figure*}

\begin{figure*}[t]
  % \hspace{-0.4cm} 
  \includegraphics[width=\linewidth]{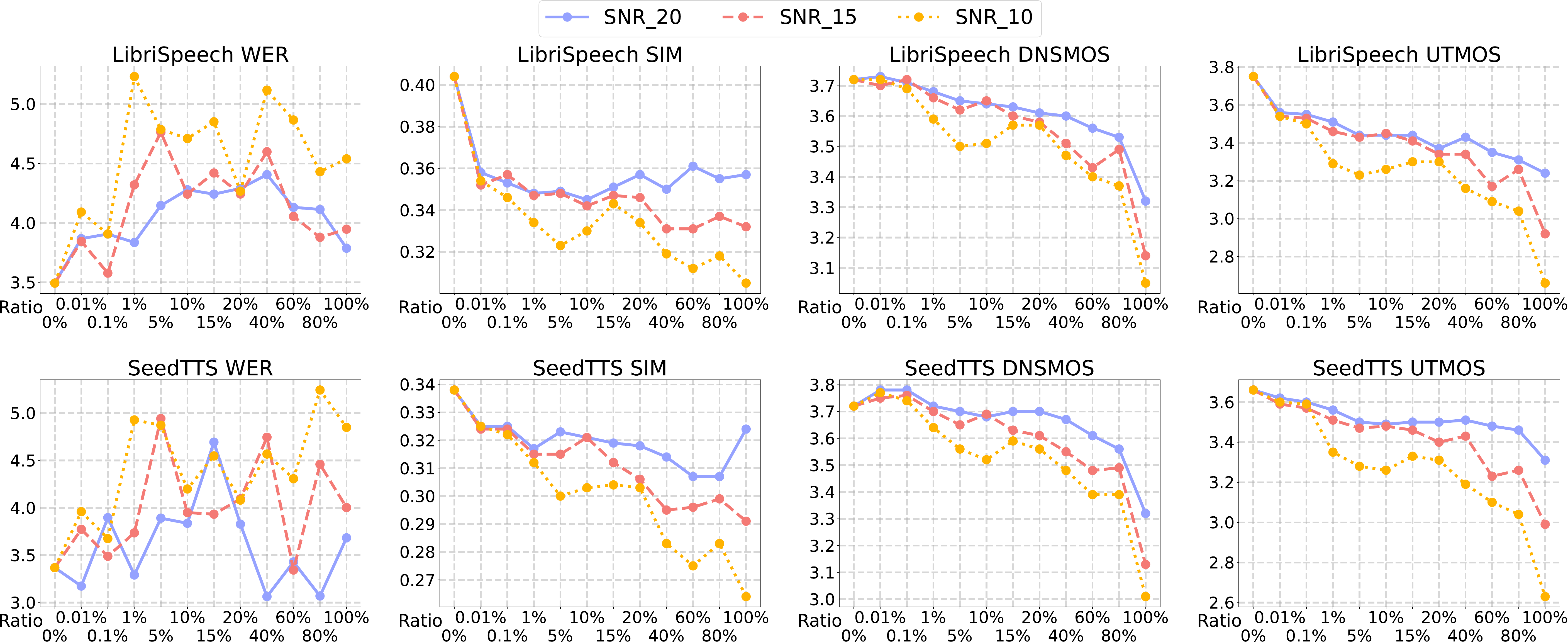}
  % \vspace{-0.2cm}
  \caption{Analysis of effects of the Noise Ratio}
  \label{fig:ratio}
  % %\vspace{-0.45cm}
\end{figure*}

\subsection{Effects of Noise Robustness }
\label{sec:noise}

\subsubsection{Noise Level} 
Noise level refers to the amount of unwanted variation or interference in a signal, which can distort or degrade the quality of the signal. Signal-to-noise ratio (SNR) is commonly used to quantify this noise level. SNR is the ratio of the power of the signal to the power of the noise and is typically expressed in decibels (dB). SNR is defined in Eq.\ref{eq:snr}, where $P_{\text{signal}}$ and $P_{\text{noise}}$ represent the power of the signal and noise, respectively.
\begin{equation}
    \text{SNR} = 10 \cdot \log_{10} \left( P_{\text{signal}}/P_{\text{noise}} \right)
    \label{eq:snr}
\end{equation}

Figure~\ref{fig:snr} presents the performance of models trained on datasets with additive noise at varying noise levels. It is observed that as the SNR decreases (i.e., noise level increases), the model performance deteriorates across all evaluation metrics, with a particularly pronounced decline when the SNR falls below 30 dB.
This phenomenon helps explain why models trained on synthetic data—typically characterized by lower background noise—can outperform those trained on real-world data, which often contain more environmental noise and interference. For example, the average DNSMOS score of the real training dataset $\mathcal{R}$ is 3.91, while that of the synthetic dataset $\mathcal{F}_{XTTS-v2}$ is higher at 3.99, confirming that the synthetic dataset has cleaner acoustic conditions with less background noise.

While it is commonly believed that cleaner data leads to better performance, this is not always the case. A plateau is observed at higher SNR values, especially in WER, SIM and UTMOS, where further increases in SNR result in marginal gains in performance. This suggests that once the cleanliness of the speech data reaches a certain threshold—where background noise is sufficiently minimized—further optimization to reduce background noise does not significantly contribute to performance improvements. On the other hand, DNSMOS consistently improves as SNR increases, primarily because it evaluates the cleanliness and naturalness of speech. This trend indicates that the quality of the generated speech is highly dependent on the characteristics of the training dataset. 

\begin{figure*}[t]
  % \hspace{-0.4cm} 
  \includegraphics[width=\linewidth]{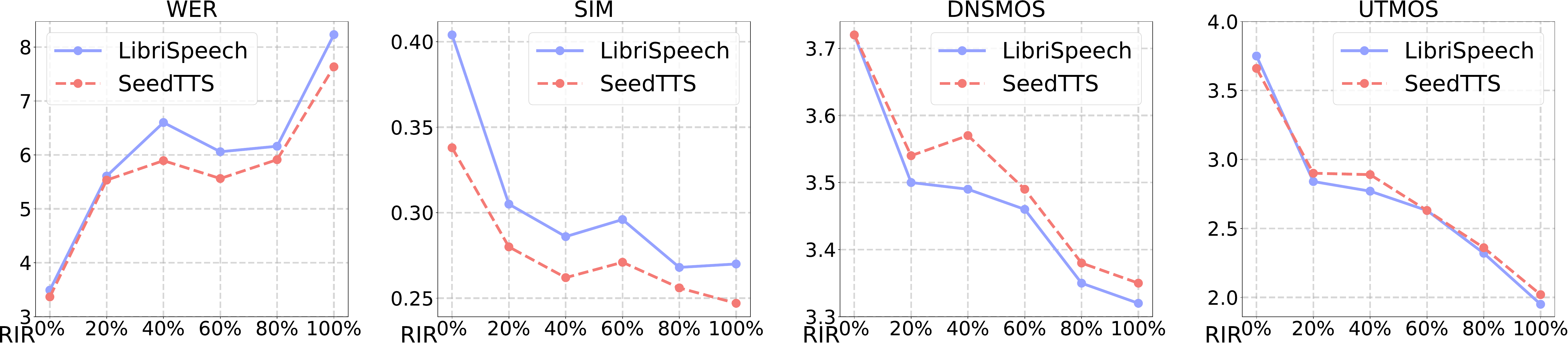}
  % \vspace{-0.2cm}
  \caption{Analysis of effects of the reverberation}
  \label{fig:rir}
  % %\vspace{-0.45cm}
\end{figure*}

% \vspace{-0.2cm}
\subsubsection{Noise Ratio}
To further investigate the impact of dataset noise on model training, we select three noise levels near the turning point (SNR=10, 15, 20). We then conduct experiments by adjusting the proportion of noisy speech within the dataset from 0\% to 100\%. As shown in Figure~\ref{fig:ratio}, the TTS model is highly sensitive to noisy data, with performance significantly degrading as the proportion of noisy speech increases. It is observed that TTS model exhibits high sensitivity to noisy data, and its impact can only be mitigated when the proportion of noisy speech is maintained at or below 0.1\%. This suggests that the contaminated noisy data can significantly degrade the model's performance,  highlighting the importance of careful data processing for TTS systems. 

% \vspace{-0.2cm}
\subsubsection{Reverberation}
% \vspace{-0.05cm}  
Besides additive noise, reverberation is another type of noise. We convolve the simulated\_rirs data (a subset of RIRS) with clean speech to simulate reverberant data. Figure~\ref{fig:rir} presents the results of the experiment where models are trained with data containing reverberation. The results indicate that the TTS model's performance is significantly affected by reverberation, similar to the presence of additive noise. The more reverberation present in the data, the worse the model's performance. The reverberation degrades the model's ability to generate clear and intelligible speech. This suggests that handling reverberation is crucial for maintaining the quality of synthesized speech, and appropriate techniques such as speech enhancement should be employed in advance to mitigate its effects in the training data. 

\subsection{Effects of Speaking Style}
% \blue{explain }
Since existing large-scale TTS models are trained on datasets with different focuses, their performance varies accordingly. Models like XTTS-v2 excel in reading-style sentences, delivering clear and well-articulated speech, whereas models such as CosyVoice2 and ChatTTS achieve a high level of naturalness, making the generated speech resemble everyday conversations. To investigate the impact of different speech styles on training, 
 We employ these three models to generate training datasets. As shown in Table \ref{tab:feasibility},  although three models trained on synthetic data achieve high DNSMOS scores, indicating a clean background and overall high quality, the TTS model trained on a more standardized synthetic dataset demonstrates superior performance, particularly in WER and SIM. This discrepancy arises because conversational speech often includes word contractions, irregular pronunciations, and informal speech patterns~\cite{zhang2024covomix}, making it more challenging for TTS models to effectively leverage these less structured, naturally styled synthetic samples. It is worth noting that model trained on dataset synthesized by CosyVoice2 performs best on DNSMOS and UTMOS. This indicates a natural but not oral dataset can guide TTS model to generate high-naturalness speeches.  

\subsection{Generalization Capability}
To evaluate the generalization capability of the TTS model, we compare its performance not only on the in-domain test set but also on the out-of-domain SeedTTS test set. As shown in Figures \ref{fig:spk}, \ref{fig:text}, \ref{fig:snr}, and \ref{fig:rir}, TTS models trained exclusively on synthetic data exhibit a slight degradation in speaker similarity (SIM) when evaluated on out-of-domain data. However, their performance in terms of Word Error Rate (WER), DNSMOS, and UTMOS remains comparable to that observed on the in-domain dataset. This suggests that while the model demonstrates a reasonable degree of generalization and robustness to unseen conditions, challenges persist when synthesizing speech for out-of-domain and previously unseen speakers.

Moreover, we observe that the model’s sensitivity to factors such as text richness, speaker diversity, and noise levels remains consistent across both in-domain and out-of-domain datasets. Importantly, the overall speech quality is largely unaffected by the domain shift, indicating the model’s robustness in maintaining stable performance across varying data domains.

\section{Conclusion and Future Work}
In this study, we systematically investigate the feasibility of training text-to-speech (TTS) models solely using synthetic data, while analyzing their sensitivity to various factors inherent to the training data. Our experimental results demonstrate that enhancing speaker and text diversity, selecting cleaner training samples, and adopting more standardized speaking styles all contribute to improved model performance.

Due to resource constraints, all experiments in this work are conducted using a single model architecture, Matcha-TTS. In future work, we plan to evaluate a broader range of TTS architectures to generalize our conclusions further. Moreover, we intend to scale up both the dataset size and model capacity to validate these findings on larger and more diverse datasets. Additionally, we aim to leverage the insights gained from this study to guide the development and training of more effective and powerful large-scale TTS models.

\section{Acknowledgment}
This work was supported in part by National Key Research and Development Program of China under Grant 2024YFC2418303, in part by the Emerging Interdisciplinary Research Specialized Project of Shanghai Municipal Health Commission (No. 2022JC024), in part by the Key Research and Development Program of Jiangsu Province, China (Grant No. BE2022059-4), and in part by SJTU Med-X (Medicine \& Engineering) Translational Research Grant (YG2025LC09) .

\bibliographystyle{IEEEtran}
\bibliography{mybib}   

% Generated by IEEEtran.bst, version: 1.14 (2015/08/26)
\begin{thebibliography}{10}
\providecommand{\url}[1]{#1}
\csname url@samestyle\endcsname
\providecommand{\newblock}{\relax}
\providecommand{\bibinfo}[2]{#2}
\providecommand{\BIBentrySTDinterwordspacing}{\spaceskip=0pt\relax}
\providecommand{\BIBentryALTinterwordstretchfactor}{4}
\providecommand{\BIBentryALTinterwordspacing}{\spaceskip=\fontdimen2\font plus
\BIBentryALTinterwordstretchfactor\fontdimen3\font minus \fontdimen4\font\relax}
\providecommand{\BIBforeignlanguage}[2]{{%
\expandafter\ifx\csname l@#1\endcsname\relax
\typeout{** WARNING: IEEEtran.bst: No hyphenation pattern has been}%
\typeout{** loaded for the language `#1'. Using the pattern for}%
\typeout{** the default language instead.}%
\else
\language=\csname l@#1\endcsname
\fi
#2}}
\providecommand{\BIBdecl}{\relax}
\BIBdecl

\bibitem{tan2021survey}
X.~Tan, T.~Qin, F.~Soong, and T.-Y. Liu, ``A survey on neural speech synthesis,'' \emph{arXiv preprint arXiv:2106.15561}, 2021.

\bibitem{lengprompttts}
Y.~Leng, Z.~Guo, K.~Shen, Z.~Ju, X.~Tan, E.~Liu, Y.~Liu, D.~Yang, K.~Song, L.~He \emph{et~al.}, ``{PromptTTS} 2: Describing and generating voices with text prompt,'' in \emph{The Twelfth International Conference on Learning Representations}.

\bibitem{zhang2024covomix}
L.~Zhang, Y.~Qian, L.~Zhou, S.~Liu, D.~Wang, X.~Wang, M.~Yousefi, Y.~Qian, J.~Li, L.~He \emph{et~al.}, ``{CoVoMix}: Advancing zero-shot speech generation for human-like multi-talker conversations,'' \emph{Proceedings of the 38th International Conference on Neural Information Processing Systems}, 2024.

\bibitem{wang2023neural}
C.~Wang, S.~Chen, Y.~Wu, Z.~Zhang, L.~Zhou, S.~Liu, Z.~Chen, Y.~Liu, H.~Wang, J.~Li \emph{et~al.}, ``Neural codec language models are zero-shot text to speech synthesizers,'' \emph{arXiv preprint arXiv:2301.02111}, 2023.

\bibitem{le2024voicebox}
M.~Le, A.~Vyas, B.~Shi, B.~Karrer, L.~Sari, R.~Moritz, M.~Williamson, V.~Manohar, Y.~Adi, J.~Mahadeokar \emph{et~al.}, ``Voicebox: Text-guided multilingual universal speech generation at scale,'' \emph{Advances in neural information processing systems}, vol.~36, 2024.

\bibitem{casanova2024xtts}
E.~Casanova, K.~Davis, E.~Gölge, G.~Göknar, I.~Gulea, L.~Hart, A.~Aljafari, J.~Meyer, R.~Morais, S.~Olayemi, and J.~Weber, ``{XTTS}: A massively multilingual zero-shot text-to-speech model,'' in \emph{Proc. {INTERSPEECH}}, 2024, pp. 4978--4982.

\bibitem{zhang2025covomix2}
L.~Zhang, Y.~Qian, X.~Wang, M.~Thakker, D.~Wang, J.~Yu, H.~Wu, Y.~Hu, J.~Li, Y.~Qian \emph{et~al.}, ``{CoVoMix2}: Advancing zero-shot dialogue generation with fully non-autoregressive flow matching,'' \emph{Proceedings of the 39th International Conference on Neural Information Processing Systems}, 2025.

\bibitem{10800670}
T.~Zhou, L.~Zhang, and Y.~Qian, ``Knowledge distillation from discriminative model to generative model with parallel architecture for speech enhancement,'' in \emph{Proc. {ISCSLP}}, 2024.

\bibitem{liu2025e2e}
Y.~Liu, Z.~Chen, L.~Zhang, and Y.~Qian, ``{E2E-BPVC}: End-to-end background-preserving voice conversion via in-context learning,'' in \emph{Proc. {INTERSPEECH}}, 2025, pp. 1378--1382.

\bibitem{guo2025splitmeanflow}
Y.~Guo, W.~Wang, Z.~Yuan, R.~Cao, K.~Chen, Z.~Chen, Y.~Huo, Y.~Zhang, Y.~Wang, S.~Liu \emph{et~al.}, ``{SplitMeanFlow}: Interval splitting consistency in few-step generative modeling,'' \emph{arXiv preprint arXiv:2507.16884}, 2025.

\bibitem{kaplan2020scaling}
J.~Kaplan, S.~McCandlish, T.~Henighan, T.~B. Brown, B.~Chess, R.~Child, S.~Gray, A.~Radford, J.~Wu, and D.~Amodei, ``Scaling laws for neural language models,'' \emph{arXiv preprint arXiv:2001.08361}, 2020.

\bibitem{li2025less}
C.~Li, W.~Zhang, W.~Wang, R.~Scheibler, K.~Saijo, S.~Cornell, Y.~Fu, M.~Sach, Z.~Ni, A.~Kumar \emph{et~al.}, ``Less is more: Data curation matters in scaling speech enhancement,'' \emph{arXiv preprint arXiv:2506.23859}, 2025.

\bibitem{zhang2024beyond}
W.~Zhang, K.~Saijo, J.-w. Jung, C.~Li, S.~Watanabe, and Y.~Qian, ``Beyond performance plateaus: A comprehensive study on scalability in speech enhancement,'' \emph{arXiv preprint arXiv:2406.04269}, 2024.

\bibitem{lajszczak2024base}
M.~{\L}ajszczak, G.~C{\'a}mbara, Y.~Li, F.~Beyhan, A.~van Korlaar, F.~Yang, A.~Joly, {\'A}.~Mart{\'\i}n-Cortinas, A.~Abbas, A.~Michalski \emph{et~al.}, ``Base tts: Lessons from building a billion-parameter text-to-speech model on 100k hours of data,'' \emph{arXiv preprint arXiv:2402.08093}, 2024.

\bibitem{anastassiou2024seed}
P.~Anastassiou, J.~Chen, J.~Chen, Y.~Chen, Z.~Chen, Z.~Chen, J.~Cong, L.~Deng, C.~Ding, L.~Gao \emph{et~al.}, ``Seed-{TTS}: A family of high-quality versatile speech generation models,'' \emph{arXiv preprint arXiv:2406.02430}, 2024.

\bibitem{bellovin2019privacy}
S.~M. Bellovin, P.~K. Dutta, and N.~Reitinger, ``Privacy and synthetic datasets,'' \emph{Stan. Tech. L. Rev.}, vol.~22, p.~1, 2019.

\bibitem{zhang2025noisepreserve}
L.~Zhang, W.~Zhang, Z.~Chen, and Y.~Qian, ``Advanced zero-shot text-to-speech for background removal and preservation with controllable masked speech prediction,'' in \emph{Proc. {ICASSP}}, 2025.

\bibitem{zhang2021denoispeech}
C.~Zhang, Y.~Ren, X.~Tan, J.~Liu, K.~Zhang, T.~Qin, S.~Zhao, and T.-Y. Liu, ``Denoispeech: Denoising text to speech with frame-level noise modeling,'' in \emph{Proc. {ICASSP}}.\hskip 1em plus 0.5em minus 0.4em\relax IEEE, 2021, pp. 7063--7067.

\bibitem{fazel2021synthasr}
A.~Fazel, W.~Yang, Y.~Liu, R.~Barra-Chicote, Y.~Meng, R.~Maas, and J.~Droppo, ``{SynthASR}: Unlocking synthetic data for speech recognition,'' in \emph{Proc. {{INTERSPEECH}}}, 2021, pp. 896--900.

\bibitem{eigenschink2023deep}
P.~Eigenschink, T.~Reutterer, S.~Vamosi, R.~Vamosi, C.~Sun, and K.~Kalcher, ``Deep generative models for synthetic data: A survey,'' \emph{IEEE Access}, vol.~11, pp. 47\,304--47\,320, 2023.

\bibitem{rossenbach2024effect}
N.~Rossenbach, B.~Hilmes, and R.~Schl{\"u}ter, ``On the effect of purely synthetic training data for different automatic speech recognition architectures,'' \emph{arXiv preprint arXiv:2407.17997}, 2024.

\bibitem{du2021synaug}
C.~Du, B.~Han, S.~Wang, Y.~Qian, and K.~Yu, ``{SynAug}: Synthesis-based data augmentation for text-dependent speaker verification,'' in \emph{Proc. ICASSP}, 2021, pp. 5844--5848.

\bibitem{zhang2024scale}
L.~Zhang, W.~Zhang, C.~Li, and Y.~Qian, ``Scale this, not that: Investigating key dataset attributes for efficient speech enhancement scaling,'' \emph{arXiv preprint arXiv:2412.14890}, 2024.

\bibitem{zhang2024ddtse}
L.~Zhang, Y.~Qian, L.~Yu, H.~Wang, H.~Yang, S.~Liu, L.~Zhou, and Y.~Qian, ``{DDTSE}: Discriminative diffusion model for target speech extraction,'' \emph{IEEE Spoken Language Technology Workshop}, 2024.

\bibitem{ueda2024exploring}
L.~H. Ueda, L.~B. Marques, F.~O. Simoes, M.~U. Neto, F.~Runstein, B.~D. B{\'o}, and P.~D. Costa, ``Exploring synthetic data for cross-speaker style transfer in style representation based tts,'' \emph{arXiv preprint arXiv:2409.17364}, 2024.

\bibitem{finkelstein2022training}
L.~Finkelstein, H.~Zen, N.~Casagrande, C.-a. Chan, Y.~Jia, T.~Kenter, A.~Petelin, J.~Shen, V.~Wan, Y.~Zhang \emph{et~al.}, ``Training text-to-speech systems from synthetic data: A practical approach for accent transfer tasks,'' \emph{arXiv preprint arXiv:2208.13183}, 2022.

\bibitem{chen2024f5}
Y.~Chen, Z.~Niu, Z.~Ma, K.~Deng, C.~Wang, J.~Zhao, K.~Yu, and X.~Chen, ``F5-tts: A fairytaler that fakes fluent and faithful speech with flow matching,'' \emph{arXiv preprint arXiv:2410.06885}, 2024.

\bibitem{ChatTTS2024}
2noise, ``Chattts,'' \url{https://github.com/2noise/ChatTTS}, 2024.

\bibitem{borsos2023soundstorm}
Z.~Borsos, M.~Sharifi, D.~Vincent, E.~Kharitonov, N.~Zeghidour, and M.~Tagliasacchi, ``Soundstorm: Efficient parallel audio generation,'' \emph{arXiv preprint arXiv:2305.09636}, 2023.

\bibitem{zhang2023speak}
Z.~Zhang, L.~Zhou, C.~Wang, S.~Chen, Y.~Wu, S.~Liu, Z.~Chen, Y.~Liu, H.~Wang, J.~Li \emph{et~al.}, ``Speak foreign languages with your own voice: Cross-lingual neural codec language modeling,'' \emph{arXiv preprint arXiv:2303.03926}, 2023.

\bibitem{wang2024speechx}
X.~Wang, M.~Thakker, Z.~Chen, N.~Kanda, S.~E. Eskimez, S.~Chen, M.~Tang, S.~Liu, J.~Li, and T.~Yoshioka, ``Speechx: Neural codec language model as a versatile speech transformer,'' \emph{IEEE/ACM Transactions on Audio, Speech, and Language Processing}, 2024.

\bibitem{du2024cosyvoice}
Z.~Du, Y.~Wang, Q.~Chen, X.~Shi, X.~Lv, T.~Zhao, Z.~Gao, Y.~Yang, C.~Gao, H.~Wang \emph{et~al.}, ``Cosyvoice 2: Scalable streaming speech synthesis with large language models,'' \emph{arXiv preprint arXiv:2412.10117}, 2024.

\bibitem{ren2020fastspeech}
Y.~Ren, C.~Hu, X.~Tan, T.~Qin, S.~Zhao, Z.~Zhao, and T.-Y. Liu, ``{FastSpeech 2}: Fast and high-quality end-to-end text to speech,'' in \emph{International Conference on Learning Representations}.

\bibitem{mehta2024matcha}
S.~Mehta, R.~Tu, J.~Beskow, {\'E}.~Sz{\'e}kely, and G.~E. Henter, ``{Matcha-TTS}: A fast tts architecture with conditional flow matching,'' in \emph{Proc. {ICASSP}}.\hskip 1em plus 0.5em minus 0.4em\relax IEEE, 2024, pp. 11\,341--11\,345.

\bibitem{hinton2015distilling}
G.~Hinton, O.~Vinyals, and J.~Dean, ``Distilling the knowledge in a neural network,'' \emph{arXiv preprint arXiv:1503.02531}, 2015.

\bibitem{zhang2021knowledge}
L.~Zhang, Z.~Chen, and Y.~Qian, ``Knowledge distillation from multi-modality to single-modality for person verification,'' \emph{Proc. Interspeech 2021}, pp. 1897--1901, 2021.

\bibitem{cheng2024residual}
J.~Cheng, R.~Liang, L.~Zhou, L.~Zhao, C.~Huang, and B.~W. Schuller, ``Residual fusion probabilistic knowledge distillation for speech enhancement,'' \emph{IEEE/ACM Transactions on Audio, Speech, and Language Processing}, 2024.

\bibitem{9414408}
M.-J. Hwang, R.~Yamamoto, E.~Song, and J.-M. Kim, ``Tts-by-tts: Tts-driven data augmentation for fast and high-quality speech synthesis,'' in \emph{Proc. {ICASSP}}, 2021, pp. 6598--6602.

\bibitem{panayotov2015librispeech}
V.~Panayotov, G.~Chen, D.~Povey, and S.~Khudanpur, ``Librispeech: an asr corpus based on public domain audio books,'' in \emph{Proc. {ICASSP}}.\hskip 1em plus 0.5em minus 0.4em\relax IEEE, 2015, pp. 5206--5210.

\bibitem{wichern2019wham}
G.~Wichern, J.~Antognini, M.~Flynn, L.~R. Zhu, E.~McQuinn, D.~Crow, E.~Manilow, and J.~L. Roux, ``Wham!: Extending speech separation to noisy environments,'' \emph{arXiv preprint arXiv:1907.01160}, 2019.

\bibitem{ko2017study}
T.~Ko, V.~Peddinti, D.~Povey, M.~L. Seltzer, and S.~Khudanpur, ``A study on data augmentation of reverberant speech for robust speech recognition,'' in \emph{Proc. {ICASSP}}.\hskip 1em plus 0.5em minus 0.4em\relax IEEE, 2017, pp. 5220--5224.

\bibitem{wang2022wespeakerresearchproductionoriented}
\BIBentryALTinterwordspacing
H.~Wang, C.~Liang, S.~Wang, Z.~Chen, B.~Zhang, X.~Xiang, Y.~Deng, and Y.~Qian, ``Wespeaker: A research and production oriented speaker embedding learning toolkit,'' 2022. [Online]. Available: \url{https://arxiv.org/abs/2210.17016}
\BIBentrySTDinterwordspacing

\bibitem{radford2023robust}
A.~Radford, J.~W. Kim, T.~Xu, G.~Brockman, C.~McLeavey, and I.~Sutskever, ``Robust speech recognition via large-scale weak supervision,'' in \emph{Proc. {ICML}}, 2023, pp. 28\,492--28\,518.

\bibitem{chen2022wavlm}
S.~Chen, C.~Wang, Z.~Chen, Y.~Wu, S.~Liu, Z.~Chen, J.~Li, N.~Kanda, T.~Yoshioka, X.~Xiao \emph{et~al.}, ``Wavlm: Large-scale self-supervised pre-training for full stack speech processing,'' \emph{IEEE Journal of Selected Topics in Signal Processing}, vol.~16, no.~6, pp. 1505--1518, 2022.

\bibitem{reddy2021dnsmos}
C.~K. Reddy, V.~Gopal, and R.~Cutler, ``{DNSMOS}: A non-intrusive perceptual objective speech quality metric to evaluate noise suppressors,'' in \emph{Proc. {ICASSP}}.\hskip 1em plus 0.5em minus 0.4em\relax IEEE, 2021, pp. 6493--6497.

\bibitem{saeki2022utmosutokyosarulabvoicemoschallenge}
T.~Saeki, D.~Xin, W.~Nakata, T.~Koriyama, S.~Takamichi, and H.~Saruwatari, ``{UTMOS}: Utokyo-sarulab system for voicemos challenge 2022,'' in \emph{Proc. {{INTERSPEECH}}}, 2022.

\end{thebibliography}

\end{document}